\documentclass[preprint]{aastex}
\usepackage{amsmath, amsthm}
\usepackage{lscape}
\usepackage{graphics}
\usepackage{epsfig}
\usepackage{graphicx}
\usepackage{rotate}
\usepackage{color}

\shorttitle{Cosmic ray tracers}
\shortauthors{Becker et al.}

\begin{document}

\title{Tracing the sources of cosmic rays with molecular ions}

\author{Julia K.\ Becker}
\affil{Ruhr-Universit\"at Bochum, Fakult\"at f\"ur Physik \&
  Astronomie, Theoretische Physik IV, D-44780 Bochum, Germany}
\email{julia@tp4.rub.de}
\author{John H.\ Black}
\affil{Dept. of Earth and Space Sciences, Chalmers University
of Technology, Onsala Space Observatory,
SE-439\ 92 Onsala, Sweden}
\author{Mohammadtaher Safarzadeh\altaffilmark{1}}
\affil{Dept. of Earth and Space Sciences, Chalmers University
of Technology, Onsala Space Observatory,
SE-439\ 92 Onsala, Sweden}
\author{Florian Schuppan}
\affil{Ruhr-Universit\"at Bochum, Fakult\"at f\"ur Physik \&
  Astronomie, Theoretische Physik IV, D-44780 Bochum, Germany}
\altaffiltext{1}{{\it Now at:} Dept. of Physics and Astronomy, Johns Hopkins
University, 3400 N. Charles Street
Baltimore, MD 21218, USA}

\begin{abstract}
The rate of ionization by cosmic rays in interstellar gas directly
associated with $\gamma$-ray emitting supernova remnants is for the
first time calculated to be several orders of magnitude larger than
the Galactic average. Analysis of ionization-induced chemistry yields
the first quantitative prediction of the astrophysical H$_{2}^{+}$
emission line spectrum, which should be detectable together with
H$_3^+$ lines. The predicted coincident observation of those emission
lines and $\gamma$-rays will help prove that supernova remnants are
sources of cosmic rays.
\end{abstract}


\section{Introduction}
One of the central questions in modern physics concerns the origin of cosmic
rays (CRs). The CRs themselves do not directly reveal their sources as
they interact with cosmic magnetic fields and particles 
on their way through the
Universe \citep{strong2007}. One method of identifying CR sources indirectly is
the search for high-energy photon or neutrino signals. When protons
of energy $E>200$~MeV
interact with photon or matter fields in the vicinity of the CR
production site, charged and neutral pions are produced, which then
emit high-energy neutrinos and photons. Those neutral
particles can be used to trace CRs, as they point back to the
sources. This approach is limited by (a) the low interaction cross 
sections that make neutrinos difficult to detect \citep{julias_review},
and (b) the competition with processes like inverse Compton 
scattering and bremsstrahlung 
\citep{schlickeiser2002}.
It is therefore crucial to use multiwavelength
approaches to identify CR sources through
correlation studies, and not to rely solely on the observation of
high-energy photons.

The illumination of molecular clouds by CRs from supernova
remnants (SNRs) was first suggested by
\cite{black_fazio1973,montmerle1979}. 
Most recently, observations performed with the Fermi Telescope have revealed high-energy emission from several regions
associated with a SNR located in close proximity to molecular clouds (MCs). In particular, emission at 0.2 to 200 GeV has
been detected towards the SNRs W51C \citep{abdo_w51c}, W28 \citep{abdo_w28}, W44
\citep{abdo_w44}, W49B \citep{abdo_w49b} and IC443 \citep{adbo_ic443}. SNRs are
one of the primary candidates for CR acceleration, as
discussed for example by \cite{biermann_prl2009}.
The observed gamma-ray signatures could therefore be interpreted as
$\pi^{0}-$decay products from high-energy CR interactions with
the MC. The data for these particular objects actually
seem to favor a hadronic scenario over a leptonic one, see
e.g.\ \citep{abdo_w44}. 
Due to the ambiguity of the signal, however,
other tracers of CRs are needed to have a final proof.

In this paper, we quantify a new method to identify SNRs
as CR sources, based on
CR-induced ionization. When CRs are accelerated in a SNR and then
injected into a nearby MC, the ionization rate in the 
ambient gas is expected to be enhanced compared with the ionization
caused by the general background of Galactic CRs.
The immediate products are hydrogen ions, H$^+$ and H$_2^+$, which 
react rapidly with H$_2$ and oxygen to form molecular ions. Ions like
H$_3^+$, OH$^+$, H$_2$O$^+$, and H$_3$O$^+$ are spectroscopically
observable at infrared and submm wavelengths and their abundances
are predictable as functions of the ionization rate
(e.g.\ \citep{black1998,montmerle2009}). 
Among the 18 molecular
cations now identified in the interstellar medium, it is the
reactive ions H$_3^+$, OH$^+$, and H$_2$O$^+$ that are expected to 
be the most specific tracers of the CR ionization rate
\citep{black1998,black2007}, and it was only during the last year that
the latter two species have been detected \citep{Benz2010_hyd_ion,
Bruderer2010_hyd2591,Gerin2010_Oions,Gupta2010OHp_Orion,Krelowski2010,
Neufeld2010_OHp,Ossenkopf2010H2Op,Schilke2010_H2Op_oprat,
Wyrowski2010_H2Op}. Analysis of OH$^+$ and H$_2$O$^+$ absorption in 
diffuse interstellar gas of low molecular fraction (H$_2$/H $<10$\%)
serves to calibrate the background ionizing frequency of interstellar
hydrogen at a level $\zeta_{\rm Gal} = (0.6 - 2.4) \cdot 10^{-16}$s$^{-1}$
 per atom \citep{Gerin2010_Oions,Neufeld2010_OHp,black_ionization_rate}.  
\section{Ionization in the direct vicinity of supernova remnants}
Here, we estimate the enhanced interstellar ionization rates 
expected from the observed fluxes of high-energy gamma rays toward
SNRs and predict the abundances of reactive molecular
ions to be expected in associated MCs. 
We show how correlations between GeV-TeV emission and 
vibrational and rotational spectra of  
molecular ions could be used to identify the sites of CR 
acceleration. Observational tests will be possible through 
astronomical spectroscopy at infrared and mm/submm wavelengths.
\noindent
CRs of kinetic energies between $E_p\sim 1$~MeV and
$\sim 1$~GeV ionize the interstellar medium, while those at lower
energy hardly escape their sources. We investigate five SNRs
located in or close to MCs, which are detected in
gamma-rays.

\subsection{Primary particle spectra}
The CR spectra, further used to calculate the ionization rate,
can be deduced from the measured gamma-ray flux at Earth:
\begin{equation}
\frac{dN_{\gamma}}{dE_{\gamma}~dt~dA_{Earth}}.
\end{equation}
Assuming that the photons arise from hadronic interactions between the
SNR and the MC, a primary CR
spectrum can be derived. We assume a smoothed broken power-law for the primaries:
\begin{equation}
\frac{dN_{p}}{dE_{p}~dt~dA_{source}} =a_p
\left(\frac{p}{1\rm{~GeV~c^{-1}}}\right)^{-s}\left(1+\left(\frac{p}{p_{br}}\right)\right)^{-\Delta
s}\,,
\end{equation}
where $a_p$ is the proton spectrum normalization and $p_{br}$ the location of the spectral break. For all sources, this
shape of the primary spectrum reproduces the shape of the observed
photon spectrum. Our results generally follow the original calculations by Fermi, \citep{abdo_w51c,abdo_w28,abdo_w44,abdo_w49b,adbo_ic443}
with slight modifications of the smoothing of the power-law, only giving minor changes to the result.

We use the parametrization of the CR interaction model as presented in \cite{kamae2006} to derive the
primary spectrum from the gamma-ray observation. We use a fixed
value of the hydrogen density, $n_H=100$ cm$^{-3}$ and determine the  CR energy $W_p$ needed
to provide the observed gamma-ray flux. The volume of the emitting region $V$ enters the normalization of the proton spectrum. It can be derived from gamma-observations and is taken from the Fermi publications.
The spectral 
parameters $s$, $\Delta s$ and $p_{br}$, the total energy budget of interacting protons, $W_p$  and the volume $V$ are listed in Table \ref{params:tab}.

\begin{table}[h!]
\centering{
\begin{tabular}{|l||l|l|l|l|l|}
\hline
&W51C&W44&W28&IC443&W49B\\\hline\hline
$s$&1.5&1.74&1.7&2.09&2.0\\\hline
$\Delta s$&1.4&1.96&1.0&0.78&1.4\\\hline
$p_{br}$ [GeV c$^{-1}$]&15&9&2&69&4\\\hline
$W_p$ [10$^{49}$ erg]&9.0&12.1&3.3&28.0&35.0\\\hline
$V$ [cm$^3$]&$3.3\cdot 10^{60}$&$4.2\cdot 10^{59}$&$3.2 \cdot 10^{59}$&$4.2\cdot 10^{59}$&$6.3\cdot 10^{56}$\\\hline
\end{tabular}
\caption{Parameters used to derive the primary cosmic ray spectrum
  from the observed photon spectrum \label{params:tab}. The total energy budget of the sources and the spectral shape are determined from the observed gamma-ray spectrum, the interacting volume is deduced from the emitting region (see the original Fermi papers).}
}
\end{table}

\subsection{Ionization}
To calculate the ionization rate, we use the parametrization
from \cite{padovani2009}, including ionization by CR
electrons, CR protons and electron capture by CR
protons. Only the direct ionization by 
primary protons turns out to be significant, the other contributions being at 
least one order of magnitude smaller.
The ionization rate of H$_2$ by primary protons reduces to 
\begin{equation}
 \zeta^{H_2} = \int_{E_{\min}}^{E_{\max}} \frac{dN_p}{dE_p}
 \sigma_{p}^{\rm ion}(E_p) dE_p\,,
\end{equation}
with $dN_p/dE_p$ as the CR spectrum at a given kinetic energy $E_p$.
The upper integration
limit is set to $E_{\max} =$ 1 GeV, as the rapidly decreasing cross
section makes contributions at higher energies negligible. 

There are two main factors affecting the ionization rate:

\begin{itemize}
\item[(1)] The CR spectrum
at Earth as measured below $\sim 1$~GeV is modified from the spectrum
at the source by transport effects of the ISM and the solar
environment. The particle spectrum at the source can be derived from
gamma-ray data up to approximately $1$~GeV, but it must be estimated
theoretically at lower energies. The simplest approach is to
extrapolate the spectrum towards lower energies, assuming that the
acceleration mechanism works the same way at MeV energies. However,
low-energy particles might follow a different acceleration scheme 
and produce a flatter or curved spectrum, see \cite{blasi2005} and
references therein. This might lead to an overestimation of
the spectrum at the lowest energies. However, it is not clear yet at
what energy this will start exactly.
\item[(2)]
The choice of lower energy threshold
$E_{\min}$ is related to propagation effects. The range of a 1 MeV 
proton in hydrogen is $8.5\times 10^{-4}$ g cm$^{-2}$, 
which corresponds to a projected column density of $5.1\times 10^{20}$ H cm$^{-2}$. This is
easily able to support detectable abundances of OH$^+$ and H$_2$O$^+$
in the diffuse molecular gas of the Galaxy. In \cite{indriolo_cr2009}, a threshold of $E_{\min} = 2$ MeV is recommended for
diffuse clouds with hydrogen number densities $n_H < 10^3~$cm$^{-3}$, and 
$E_{\min} = 10$~MeV for dense clouds, where $n_H > 10^3~$cm$^{-3}$. 
Close to a CR accelerator, at an average density of $100$~cm$^{-3}$, values above 2~MeV 
are reasonable choices to account for
the stopping range.
\end{itemize}
We chose the following approach to account for these uncertainties:
we first assume that the spectrum continues as a power-law down to MeV
energies. 
We then introduce an artificial and sharp cutoff in the spectrum at a
minimum kinetic energy
$E_{\min}$, below which the CR flux is set to zero. 
We then vary the minimum energy between
$E_{\min}=1$~MeV and $100$~MeV. As the CR ionization
cross section decreases with energy, increasing the threshold will remove much of the
potential signal, having an even more extreme effect than flattening
the CR spectrum. Increasing the threshold at the same time accounts
for considering different stopping ranges. At this stage, due to the
lack of knowledge of the lower part of the spectrum, we will calculate
the possible range of ionization intensity rather than one specific value.

\begin{figure}[ht]
 \centering
   \includegraphics[width=10cm]{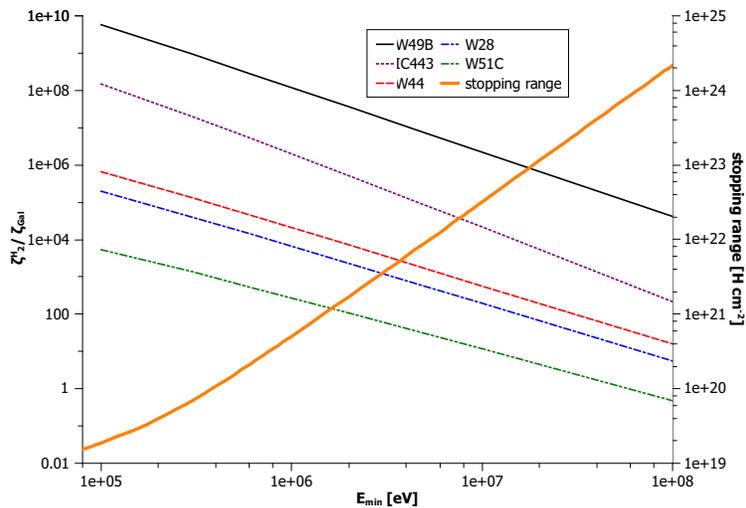} 
 \caption{ \label{fig:all}
Dependence of the ionization rate, normalized to the average Galactic value of
  $\zeta_{\rm Gal}:=2\cdot 10^{-16}$~s$^{-1}$, on $E_{\min}$. }
\end{figure}

Figure \ref{fig:all} shows the ionization rate, normalized to Galactic average, versus the minimum energy. For comparison, the stopping range
is shown as a function of $E_{\min}$.
For the most optimistic scenario, still compatible with current knowledge, $E_{\min}=2$~MeV,
ionization rates range from $110$ to $4 \cdot 10^{7}$ times the Galactic
average. An energy cutoff at $100$~MeV 
 leaves W49B as an
outstanding source with a factor of $4 \cdot 10^{4}$ above Galactic
average. The central reason is the large CR energy density in this
source \citep{abdo_w49b}.  At $E_{\min}=100$~MeV, only W51C drops to
values below $\zeta_{\rm Gal}$. All other sources remain a factor of a
few to ten above the background.

Note that we do not include the  
additional ionization by energetic secondary electrons being 
produced in the primary ionizing event. These secondary ionizations  
increase the total rate by factors of a few (cf. Padovani et al.  
2009), but the details are sensitive to the relative abundances of H,  
H$_2$, e$^-$, and He \citep{dalgarno1999}. The escape of  
CRs from their sources and their propagation into surrounding  
material are also affected in complicated ways by interaction with  
magnetic fields (see \cite{padovani_galli2011} for a recent discussion  
for dense molecular gas). These issues will be addressed in future work.

\section{Cosmic ray induced molecular ions}
As shown below, the observable line intensities of molecular ions 
spawned by CR ionization depend upon the ionization rate integrated
through a column of molecular gas
$$\eta = \int_0^L \zeta^{\rm H_2} n({\rm H}_2) d\ell \approx  \zeta^{\rm H_2}
n({\rm H}_2) L \;\;\;{\rm s}^{-1}\;{\rm cm}^{-2}\,,$$
which we call the {\it cosmic ray exposure}. The number density of 
interstellar H$_2$ is $n({\rm H}_2)$ and the integration is over path
length $\ell=0$ to $L$. Because the CRs of shortest
range die off at small $\ell$, protons at energies significantly 
greater than $E_{\rm min}$ contribute to the exposure. 
In principle, the exposure 
should be computed through an explicit depth-integration of the discrete 
energy loss. 
The SNRs discussed in this context have a CR exposure of the
order of $\eta \sim 10^9$ s$^{-1}$ cm$^{-2}$ or higher (see Fig.\ \ref{fig:all}), which 
is $> 10^4$ times larger than the exposures that characterize diffuse
clouds in the Galactic disk.
In general, ionization by X-rays can produce similar chemical signatures.
We expect the X-rays from SNRs to be much less important than CRs because the X-ray luminosities are relatively small and their
depths of penetration are too short to yield high values of the exposure.

The outlines of interstellar ion chemistry
have been understood for almost 40 years \citep{Herbst_Klemperer1973} and 
the rate coefficients of all the crucial reactions are well
determined. In steady state between the rates of source and sink reactions,
the number densities of the transient ions O$^+$, OH$^+$, H$_2$O$^+$, and
HeH$^+$ are expected to be proportional to $\zeta_{\rm H}$. In fully molecular 
regions, $n({\rm H})<<n({\rm H}_2)$, of low ionization $x(e^-)<10^{-5}$, 
competing processes rarely interrupt the sequence of H-atom abstraction
reactions so that almost every ionization of hydrogen leads to formation of
H$_2^+$, H$_3^+$, 
OH$^+$, H$_2$O$^+$, and H$_3$O$^+$. The time-scales to achieve 
steady state are very short, $\sim 1 \bigl(1000\;{\rm cm}^{-3}/n({\rm H}_2)
\bigr)$ year. 

The abundances of interstellar molecules can be computed from a system
of rate equations for various values of the physical conditions (density,
temperature, ionization rate) and gas-phase elemental abundances, either
in the steady-state limit or as a time-dependent solution. However, the
abundances alone do not characterize observable signatures,
because the distributions of the molecules among their quantum states
tend to be far out of equilibrium. The departures from equilibrium will
be most extreme for the most reactive molecular ions, because elastic 
and inelastic collisions that would otherwise thermalize them are 
typically not faster than the reactive collisions with the main collision
partner, H$_2$, that destroy them \citep{black1998}. We illustrate the
coupled effects of ion chemistry and non-equilibrium excitation with
results from a reference model of molecular gas at density $n({\rm H}_2)
=100$ cm$^{-3}$ exposed to an ionization rate $\zeta^{\rm H_2}=10^{-12}$~s$^{-1}$
over a path length $L=10^{19}$~cm~$=3.24$ pc. The excitation and
transport of spectrum lines are computed through use of the {\tt Radex}
program, which can incorporate formation and destruction rates 
state-by-state for reactive molecules along with all relevant radiative
and inelastic collision processes \citep{Radex2007}. The kinetic temperature of the molecular gas  
is assumed to be 150 K, somewhat elevated owing to enhanced heating  
by cosmic rays. All line widths are taken to be 9.394 km s$^{-1}$.
 Full details of these
calculations will be presented elsewhere.  
Because H$_2^+$ has no permanent dipole moment, its vibration-rotation
and pure rotational spectrum occurs only as weak electric quadrupole transitions. Upon birth by ionization of H$_2$, H$_2^+$ is endowed
with high vibrational excitation but it retains the relatively cold
rotational distribution of its parent. At gas densities below 
$10^4$ cm$^{-3}$, excited H$_2^+$ decays to the ground state by radiation
more rapidly than it is destroyed or excited by collisions. As a result,
the emitted spectrum in our reference model is directly predictable 
from the value of the CR exposure, $\eta=10^9$ s$^{-1}$ cm$^{-2}$,
and the known quadrupole transition probabilities \citep{posen1983}. Owing 
to symmetry-breaking in the last few bound vibration-rotation levels near
the dissociation limit, H$_2^+$ exhibits a few dipole-allowed 
radio-frequency transitions \citep{brown-carrington-book,howells1991,
moss1993,bunker2000,critchley2001}. These 
transitions are excited in our model by the small fraction of ionizations
that put H$_2^+$ into levels $(v=19,J=0,1)$ of the 1s$\sigma_{\rm g}$
ground electronic state. Transitions to three barely bound levels $(v=0,
J=0,1,2)$ of the 2p$\sigma_{\rm u}$ state occur with high probabilities
\citep{moss1993}
that partly compensate for the low populations of the initial states: 
we predict that these transitions at 17.6, 52.9, and 96.4 GHz will 
appear as weak masers in regions of high CR exposure. The
electronically excited levels also radiate rapidly to 1s$\sigma_{\rm g}
(v=17-18,J=0-3)$ in transitions at mm/submm wavelengths. Although these 
symmetry-breaking transitions will probably be undetectably weak in 
sources like those considered here, they may become observable in 
the more extreme environments in active galactic nuclei.

\begin{figure}[ht]
 \centering
   \includegraphics[width=10cm,angle=270]{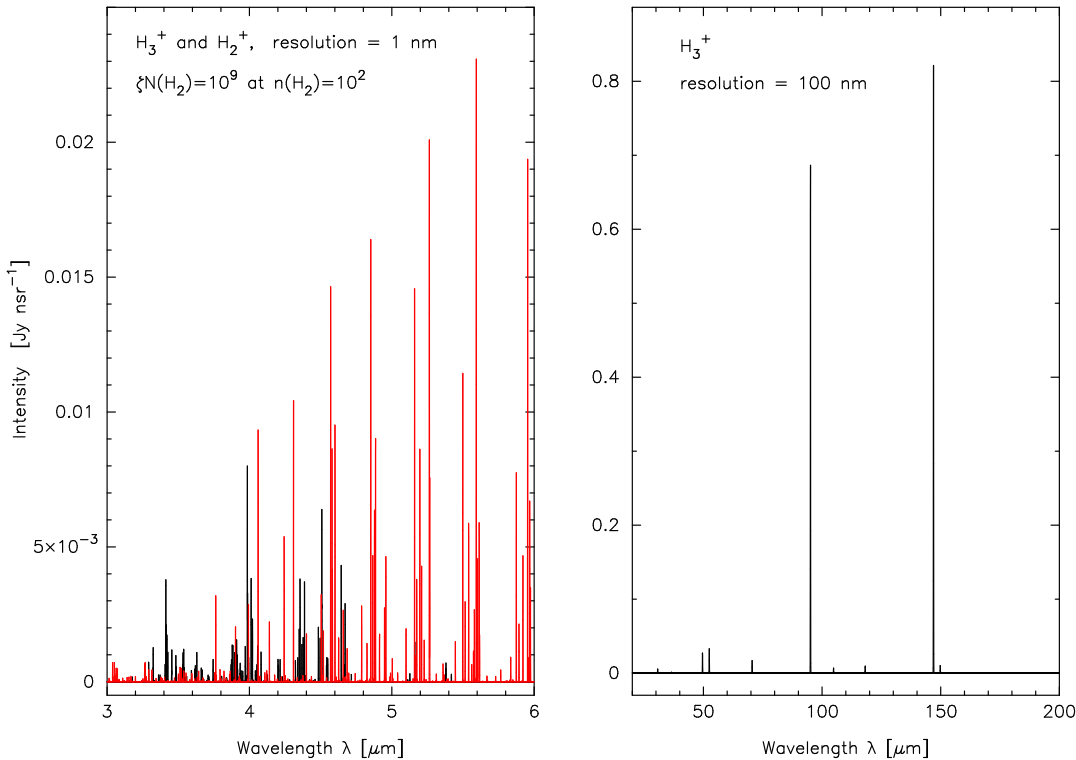}
 \caption{ \label{fig:H2+spectrum}
Predicted emission spectra of H$_2^+$ (red in online version)  
and H$_3^+$ (black online) for the reference model. The surface  
brightness is displayed in Jy nsr$^{-1}$ where 1 Jy = $10^{-26}$ W m$^ 
{-2}$ Hz$^{-1}$ and 1 nsr = $10^{-9}$ steradian of solid angle. In the
 figure, $\eta=10^{9}$~s$^{-1}$~cm$^{-2}$ is used as a typical value
 in this context (see
 text).}
\end{figure}

The high excitation of H$_2^+$ is partly transferred to H$_3^+$ in the
next chemical step and H$_3^+$ can also be excited by collisions
several times during its lifetime. The distortion-induced rotational transition \citep{pan_oka1986} of H$_3^+$ $(J,K)=(4,4)\to (3,1)$
easily suffers population inversion \citep{black1998}, a necessary condition
for a maser. The transition frequency is poorly
known (approximately 217.7 GHz), but the line is predicted in our reference
model to appear as a weak maser with a peak Rayleigh-Jeans brightness
temperature of 10 mK, an intensity that is detectable with existing 
radio telescopes. The intrinsically weak, far-infrared rotational 
transitions of H$_3^+$ are excited mainly by collisions at the kinetic
temperature of the gas, while the infrared vibration-rotation transitions
of both H$_2^+$ and H$_3^+$ are excited in the formation process with
superthermal intensity distributions. In  
general, the rotational line intensities of H$_3^+$ will depend upon  
the local density and temperature; however, the entire spectrum of H$_2^+$ is rather insensitive to these physical conditions.
We calculate explicitly the formation
rate of HeH$^+$ by vibrationally excited H$_2^+$ and predict the spectrum
that results from its initial, superthermal excitation. The predicted
spectra are illustrated in Figure \ref{fig:H2+spectrum}. 
At infrared 
wavelengths the strongest lines of H$_3^+$ appear between 3 and 4.6 $\mu$m
while H$_2^+$ lines of comparable intensity appear between 4 and 6 $\mu$m. 
The strongest vibration-rotation lines of HeH$^+$ in this region are 
several hundred times weaker. In the far-infrared region, the thermally
excited lines of H$_3^+$ are quite strong while the lines of HeH$^+$
are again considerably weaker. In the far-infrared, thermal emission by
interstellar dust particles will produce continuous radiation
with typical surface brightness $\geq 0.1$ Jy nsr$^{-1}$ at
a wavelength of 100 $\mu$m; even so, it appears that the strongest
rotational lines of H$_3^+$ might be observable if the spectroscopic
resolving power is at least 100 (resolution of 1 $\mu$m at 100 $\mu$m
wavelength). For  
comparison, the weak H$_3^+$ feature at 30.725 $\mu$m has an  
intensity approximately 1/2 as large as that of the H$_2$ $J=2\to 0$  
line at 28.226 $\mu$m. We have also calculated the spectra of the OH$_n^+$ ions, $n=1-3$.
These ions form mainly as a result of reactions of thermalized H$_3^+$ 
or H$^+$ with oxygen and subsequent reactions with 
H$_2$. The excess energy (enthalpies of reactants minus products) of
each reaction will go partly into vibrational and rotational excitation
of the product, so that the OH$_n^+$ ions will also exhibit superthermal excitation. 
A predicted signal of enhanced production of oxygen-bearing 
ions will be strong absorption in rotation-inversion transitions arising
in highly metastable states of H$_3$O$^+$ at frequencies 1.6 to 1.9 THz.
Absorption spectroscopy of interstellar H$_3^+$ has already suggested
enhanced rates of ionization throughout the central molecular zone of
the Galaxy \citep{goto2008,geballe_oka2010} and in two locations near the SNR IC~443 \citep{indriolo_ic443}. We predict that emission lines of 
both H$_2^+$ and H$_3^+$ will be detectable and that their intensities
can be used to measure the CR exposure.

\section{Summary}

We have calculated the associated cosmic ray ionization rate in interstellar matter near SNRs.
While the exact value will depend on the exact density and spectral shape, we expect ionization to be significantly above Galactic average. 
The cosmic ray exposure of the interstellar hydrogen
(ionization rate times gas density integrated over path length) for 
the SNRs W51C, W28, W44, IC443 and W49B is expected to be of the order of $\eta=10^{9}$~s$^{-1}$~cm$^{-2}$. 
This permits the construction of a coupled
model of the ion chemistry and the excitation of reactive molecular ions
that might serve as specific tracers of cosmic rays near their sources. 
Predictions are made of the intensities of rotational and 
vibration-rotation transitions of the reactive ions including the
first quantitative prediction of an astronomical spectrum of H$_2^+$.
It appears that superthermally excited emission lines of H$_2^+$ and
H$_3^+$ may be detectable at wavelengths between 3 and 6 $\mu$m. Strong
thermal emission of H$_3^+$ may be measurable in the far infrared at
50 to 150 $\mu$m wavelength. W49B seems to be an outstanding candidate
for the detection, with  predicted ionization rate
more than 4 orders of magnitude above Galactic average. Future measurements by e.g.\ Herschel and ALMA
provide the opportunity to make observations of ionization signatures in SNR/MC systems and analyze their
correlation to the gamma-ray emission. A recent observation suggests an  
enhanced excitation and abundance of ammonia toward W28 \citep{nicholas2011}. Using {\tt Radex}, it is possible to explain the observed spectrum as CR induced 
with an ionization rate of $4.5\cdot 10^{-12}$~s$^{-1}$, so a factor of $>10^{4}$ above Galactic average. Further, H$_{3}^{+}$ emission observed from the direction of IC443 indicates an enhanced ionization level of $2\cdot 10^{-15}$~s$^{-1}$, so a factor of 10 above Galactic average \citep{indriolo_ic443_h3p_2010}.
In the future, more observations need to be performed in order to explicitly search for the expected signatures of H$_2^+$ and H$_3^+$. Treating the propagation of the CRs in detail will allow us to give even more precise predictions of the expected emission regions. 

\acknowledgements{
We thank S.\ Casanova for helpful discussions,
 acknowledge support from the Research Department of
Plasmas with Complex Interactions (Bochum) and from the Lorentz
Center, Leiden, The Netherlands. Research on interstellar chemistry
at Chalmers has been supported by the Swedish Research Council and the Swedish
National Space Board.}
 

\begin{thebibliography}{44}
\providecommand{\natexlab}[1]{#1}
\providecommand{\url}[1]{\texttt{#1}}
\expandafter\ifx\csname urlstyle\endcsname\relax
  \providecommand{\doi}[1]{doi: #1}\else
  \providecommand{\doi}{doi: \begingroup \urlstyle{rm}\Url}\fi

\bibitem[{Abdo} et~al.(2009){Abdo}, {(Fermi Coll.)}, et~al.]{abdo_w51c}
A.~A. {Abdo}, {(Fermi Coll.)}, et~al.
\newblock \emph{ApJL}, 706:\penalty0 L1, 2009.

\bibitem[{Abdo} et~al.(2010{\natexlab{a}}){Abdo}, {(Fermi Coll.)},
  et~al.]{abdo_w28}
A.~A. {Abdo}, {(Fermi Coll.)}, et~al.
\newblock \emph{ApJ}, 718:\penalty0 348, 2010{\natexlab{a}}.

\bibitem[{Abdo} et~al.(2010{\natexlab{b}}){Abdo}, {(Fermi Coll.)},
  et~al.]{abdo_w44}
A.~A. {Abdo}, {(Fermi Coll.)}, et~al.
\newblock \emph{Science}, 327:\penalty0 1103, 2010{\natexlab{b}}.

\bibitem[{Abdo} et~al.(2010{\natexlab{c}}){Abdo}, {(Fermi Coll.)},
  et~al.]{abdo_w49b}
A.~A. {Abdo}, {(Fermi Coll.)}, et~al.
\newblock \emph{ApJ}, 722:\penalty0 1303, 2010{\natexlab{c}}.

\bibitem[{Abdo} et~al.(2010{\natexlab{d}}){Abdo}, {(Fermi Coll.)},
  et~al.]{adbo_ic443}
A.~A. {Abdo}, {(Fermi Coll.)}, et~al.
\newblock \emph{ApJ}, 712:\penalty0 459, 2010{\natexlab{d}}.

\bibitem[{Becker}(2008)]{julias_review}
J.~K. {Becker}.
\newblock \emph{Phys.\ Rep.}, 458:\penalty0 173, 2008.

\bibitem[{Benz} et~al.(2010)]{Benz2010_hyd_ion}
A.~O. {Benz} et~al.
\newblock \emph{A \& A}, 521:\penalty0 L35, 2010.

\bibitem[{Biermann} et~al.(2009)]{biermann_prl2009}
P.~L. {Biermann} et~al.
\newblock \emph{PRL}, 103\penalty0 (6):\penalty0 061101, 2009.

\bibitem[{Black}(1998)]{black1998}
J.~H. {Black}.
\newblock \emph{Ch.\ \& Ph.\ Mol.\ \& Gr.\ in Sp.}, 109:\penalty0 257, 1998.

\bibitem[{Black}(2007)]{black2007}
J.~H. {Black}.
\newblock In \emph{Molecules in Space and Laboratory}, 2007.

\bibitem[{Black} and {Fazio}(1973)]{black_fazio1973}
J.~H. {Black} and G.~G. {Fazio}.
\newblock \emph{ApJL}, 185:\penalty0 L7, 1973.

\bibitem[{Blasi} et~al.(2005)]{blasi2005}
P.~{Blasi} et~al.
\newblock \emph{MNRAS}, 361:\penalty0 907, 2005.

\bibitem[{Brown} and {Carrington}(2003)]{brown-carrington-book}
J.~M. {Brown} and A.~{Carrington}.
\newblock \emph{{Rotational Spectroscopy of Diatomic Molecules}}.
\newblock Cambridge Univ. Press, 2003.

\bibitem[{Bruderer} et~al.(2010)]{Bruderer2010_hyd2591}
S.~{Bruderer} et~al.
\newblock \emph{A \& A}, 521:\penalty0 L44, 2010.

\bibitem[Bunker and Moss(2000)]{bunker2000}
P.~R. Bunker and R.~E. Moss.
\newblock \emph{CPL}, 316\penalty0 (3-4):\penalty0 266, 2000.

\bibitem[{Critchley} et~al.(2001)]{critchley2001}
A.~D. {Critchley} et~al.
\newblock \emph{PRL}, 86:\penalty0 1725, 2001.

\bibitem[{Dalgarno} et~al.(1999){Dalgarno}, {Yan}, and {Liu}]{dalgarno1999}
A.~{Dalgarno}, M.~{Yan}, and W.~{Liu}.
\newblock \emph{ApJ Suppl.~Series}, 125:\penalty0 237, 1999.

\bibitem[{Geballe} and {Oka}(2010)]{geballe_oka2010}
T.~R. {Geballe} and T.~{Oka}.
\newblock \emph{ApJL}, 709:\penalty0 L70, 2010.

\bibitem[{Gerin} et~al.(2010)]{Gerin2010_Oions}
M.~{Gerin} et~al.
\newblock \emph{A \& A}, 518:\penalty0 L110, 2010.

\bibitem[{Goto} et~al.(2008)]{goto2008}
M.~{Goto} et~al.
\newblock \emph{ApJ}, 688:\penalty0 306, 2008.

\bibitem[{Gupta} et~al.(2010)]{Gupta2010OHp_Orion}
H.~{Gupta} et~al.
\newblock \emph{A \& A}, 521:\penalty0 L47, 2010.

\bibitem[{Herbst} and {Klemperer}(1973)]{Herbst_Klemperer1973}
E.~{Herbst} and W.~{Klemperer}.
\newblock \emph{ApJ}, 185:\penalty0 505, 1973.

\bibitem[{Howells} and {Kennedy}(1991)]{howells1991}
M.~H. {Howells} and R.~A. {Kennedy}.
\newblock \emph{CPL}, 184:\penalty0 521, 1991.

\bibitem[{Indriolo} et~al.(2009)]{indriolo_cr2009}
N.~{Indriolo} et~al.
\newblock \emph{ApJ}, 694:\penalty0 257, 2009.

\bibitem[{Indriolo} et~al.(2010{\natexlab{a}})]{indriolo_ic443}
N.~{Indriolo} et~al.
\newblock \emph{ApJ}, 724:\penalty0 1357, 2010{\natexlab{a}}.

\bibitem[{Indriolo} et~al.(2010{\natexlab{b}})]{indriolo_ic443_h3p_2010}
N.~{Indriolo} et~al.
\newblock \emph{ApJ}, 724:\penalty0 1357, 2010{\natexlab{b}}.

\bibitem[{Kamae} et~al.(2006)]{kamae2006}
T.~{Kamae} et~al.
\newblock \emph{ApJ}, 647:\penalty0 692, 2006.

\bibitem[{Kre{\l}owski} et~al.(2010)]{Krelowski2010}
J.~{Kre{\l}owski} et~al.
\newblock \emph{ApJL}, 719:\penalty0 L20, 2010.

\bibitem[{Montmerle}(1979)]{montmerle1979}
T.~{Montmerle}.
\newblock \emph{ApJ}, 231:\penalty0 95, 1979.

\bibitem[{Montmerle}(2009)]{montmerle2009}
T.~{Montmerle}.
\newblock \emph{ArXiv:0909.0222}, 2009.

\bibitem[{Moss}(1993)]{moss1993}
R.~E. {Moss}.
\newblock \emph{CPL}, 206:\penalty0 83, 1993.

\bibitem[{Neufeld} et~al.(2010{\natexlab{a}})]{Neufeld2010_OHp}
D.~A. {Neufeld} et~al.
\newblock \emph{A \& A}, 521:\penalty0 L10, 2010{\natexlab{a}}.

\bibitem[{Neufeld} et~al.(2010{\natexlab{b}})]{black_ionization_rate}
D.~A. {Neufeld} et~al.
\newblock \emph{A \& A}, 521:\penalty0 L10, 2010{\natexlab{b}}.

\bibitem[{Nicholas} et~al.(2011)]{nicholas2011}
B.~{Nicholas} et~al.
\newblock \emph{MNRAS}, 411:\penalty0 1367, 2011.

\bibitem[{Ossenkopf} et~al.(2010)]{Ossenkopf2010H2Op}
V.~{Ossenkopf} et~al.
\newblock \emph{A \& A}, 518:\penalty0 L111, 2010.

\bibitem[{Padovani} and {Galli}(2011)]{padovani_galli2011}
M.~{Padovani} and D.~{Galli}.
\newblock \emph{A \& A}, 530:\penalty0 A109, 2011.

\bibitem[{Padovani} et~al.(2009)]{padovani2009}
M.~{Padovani} et~al.
\newblock \emph{A \& A}, 501:\penalty0 619, 2009.

\bibitem[{Pan} and {Oka}(1986)]{pan_oka1986}
{F.-S.} {Pan} and T.~{Oka}.
\newblock \emph{ApJ}, 305:\penalty0 518, 1986.

\bibitem[{Posen} et~al.(1983)]{posen1983}
A.~G. {Posen} et~al.
\newblock \emph{Atom D \& Nucl D Tab}, 28:\penalty0 265, 1983.

\bibitem[{Schilke} et~al.(2010)]{Schilke2010_H2Op_oprat}
P.~{Schilke} et~al.
\newblock \emph{A \& A}, 521:\penalty0 L11, 2010.

\bibitem[{Schlickeiser}(2002)]{schlickeiser2002}
R.~{Schlickeiser}.
\newblock \emph{{Cosmic Ray Astrophysics}}.
\newblock Springer, 2002.

\bibitem[{Strong} et~al.(2007)]{strong2007}
A.~W. {Strong} et~al.
\newblock \emph{Ann.\ Rev.\ N.\ Part.\ Sc.}, 57:\penalty0 285, 2007.

\bibitem[{van der Tak} et~al.(2007)]{Radex2007}
F.~F.~S. {van der Tak} et~al.
\newblock \emph{A \& A}, 468:\penalty0 627, 2007.

\bibitem[{Wyrowski} et~al.(2010)]{Wyrowski2010_H2Op}
F.~{Wyrowski} et~al.
\newblock \emph{A \& A}, 521:\penalty0 L34, 2010.

\end{thebibliography}
\end{document}